\documentclass[%
reprint,
prl,
superscriptaddress,
nobibnotes,
nofootinbib,
amsmath,amssymb,
floatfix,
floats,
twocolumn,
longbibliography
]{revtex4-1}
\usepackage{graphicx}
\usepackage[utf8]{inputenc}
\usepackage[T1]{fontenc}
\usepackage{lmodern}
\usepackage[english]{babel}
\usepackage[separate-uncertainty = true,multi-part-units=single,per-mode = symbol]{siunitx}
\DeclareSIUnit\molar{M}
\begin{document}

\title{Measuring structural parameters of crosslinked and entangled semiflexible polymer networks with single-filament tracing}

\author{Tina Händler}
\affiliation{Peter Debye Institute for Soft Matter Physics, Leipzig University, Linnéstraße 5, 04103 Leipzig, Germany}
\affiliation{Fraunhofer Institute for Cell Therapy and Immunology, Perlickstraße 1, 04103 Leipzig, Germany}
\author{Cary Tutmarc}
\affiliation{Peter Debye Institute for Soft Matter Physics, Leipzig University, Linnéstraße 5, 04103 Leipzig, Germany}
\affiliation{Fraunhofer Institute for Cell Therapy and Immunology, Perlickstraße 1, 04103 Leipzig, Germany}
\author{Martin Glaser}
\affiliation{Peter Debye Institute for Soft Matter Physics, Leipzig University, Linnéstraße 5, 04103 Leipzig, Germany}
\affiliation{Fraunhofer Institute for Cell Therapy and Immunology, Perlickstraße 1, 04103 Leipzig, Germany}
\author{Jessica S. Freitag}
\affiliation{Fraunhofer Institute for Cell Therapy and Immunology, Perlickstraße 1, 04103 Leipzig, Germany}
\author{David M. Smith}
\affiliation{Peter Debye Institute for Soft Matter Physics, Leipzig University, Linnéstraße 5, 04103 Leipzig, Germany}
\affiliation{Fraunhofer Institute for Cell Therapy and Immunology, Perlickstraße 1, 04103 Leipzig, Germany}
\affiliation{Institute of Clinical Immunology, University of Leipzig Medical Faculty, 04103 Leipzig, Germany}
\affiliation{Dhirubhai Ambani Institute of Information and Communication Technology, Gandhinagar 382 007, India}
\author{Jörg Schnauß}
\affiliation{Peter Debye Institute for Soft Matter Physics, Leipzig University, Linnéstraße 5, 04103 Leipzig, Germany}
\affiliation{Fraunhofer Institute for Cell Therapy and Immunology, Perlickstraße 1, 04103 Leipzig, Germany}
\affiliation{Unconventional Computing Laboratory, Department of Computer Science, University of the West of England, Bristol, UK} 
\date{\today}

\begin{abstract}
\textbf{Single-filament tracing has been a valuable tool to directly determine geometrical and mechanical properties of entangled polymer networks. However, systematically verifying how the stiffness of the tracer filament or its molecular interactions with the surrounding network impacts the measurement of these parameters has not been possible with the established experimental systems. Here, we use mechanically programmable DNA nanotubes embedded in crosslinked and entangled F-actin networks, as well as in synthetic DNA networks, in order to measure fundamental, structural network properties like tube width and mesh size with respect to the stiffness of the tracers. While we confirm some predictions derived from models based purely on steric interactions, our results indicate that these models should be expanded to account for additional inter-filament interactions, thus describing the behavior of real polymer networks.}
\end{abstract}


\maketitle

\begin{figure}
\includegraphics[width=\columnwidth]{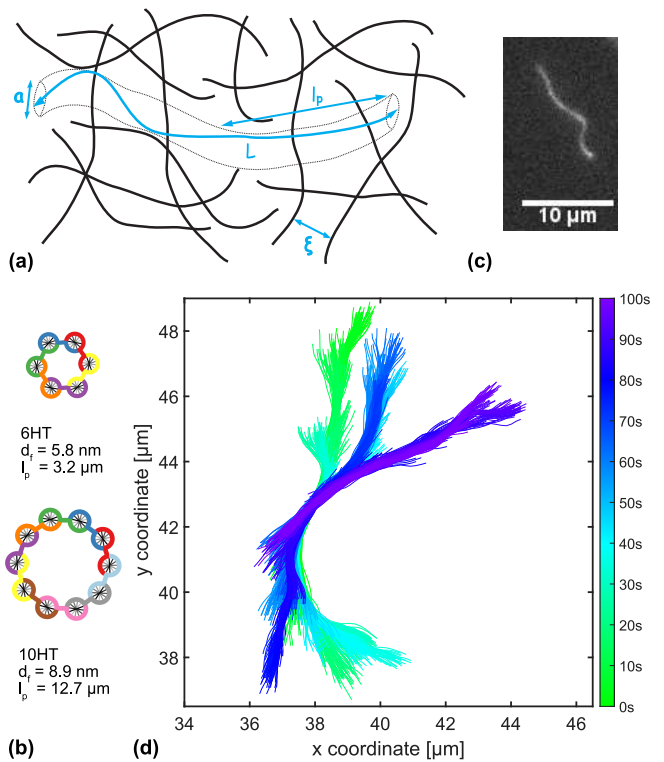}
		\caption{(a) Scheme of a tracer filament of persistence length $l_{\text{p}}$ and contour length $L$, confined to a reptation tube of width $a$ in a network of mesh size $\xi$. (b) Symbolic cross-sections of the \(n\)HT tracers with the smallest (6HT) and largest (10HT) circumference used in this study, where each color corresponds to one of the 6 or 10 oligonucleotides (adapted from \cite{schuldt_tuning_2016}). Filament diameters $d_{\text{f}}$ and persistence lengths $l_{\text{p}}$ of the other tracer filaments are in between these limits \cite{supplemental}. (c) Microscopy image of a fluorescently labeled 8HT filament. (d) Exemplary overlay of 1000 tracked frames of a reptating 8HT filament embedded in a crosslinked F-actin network (actin at \SI{0.5}{\milli\gram\per\milli\litre} with crosslinker wLX). The color coding indicates the measurement time (see color map on the side). The filament is neither crosslinked nor constrained to the original reptation tube, but able to explore the surrounding network during the observation time of \SI{100}{\second}.}
		\label{fig_fig1}
\end{figure}

\begin{figure}
  \includegraphics[width=\columnwidth]{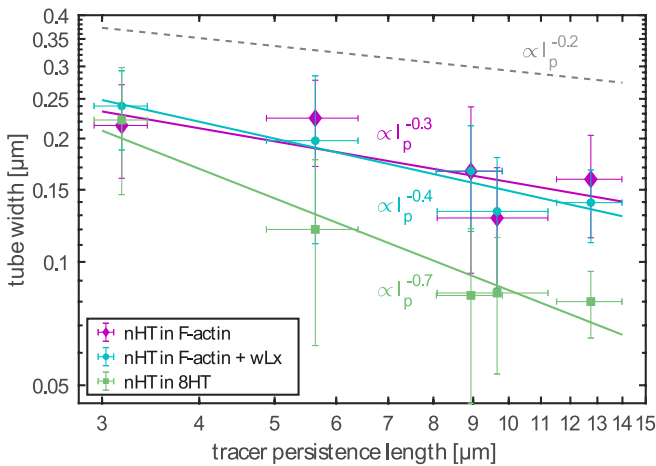}
  \caption{Using DNA nanotube tracers, we were able to examine the dependency of tube width on persistence length for three different background networks: entangled F-actin  networks (purple diamonds), crosslinked F-actin networks (crosslinked at a ratio of actin monomers to wLX of 150:1 \cite{lorenz_synthetic_2018}, blue circles) and 8HT networks (green squares). Tube width error bars are weighted standard errors obtained by calculating weighted averages of several individual tracer filaments. The tracers were 6HT, 7HT, 8HT, 9HT and 10HT, with a persistence length range from approximately \SI{3}{\micro\meter} to \SI{13}{\micro\meter} \cite{schuldt_tuning_2016, supplemental}. Matching colored lines depict the power laws resulting from fitting the leading term of Eq. \eqref{eq_xi_hinsch}. The dashed line indicates the theoretical scaling derived from the tube model.}
  \label{fig_fig2}
\end{figure}

\begin{figure}
  \includegraphics[width=\columnwidth]{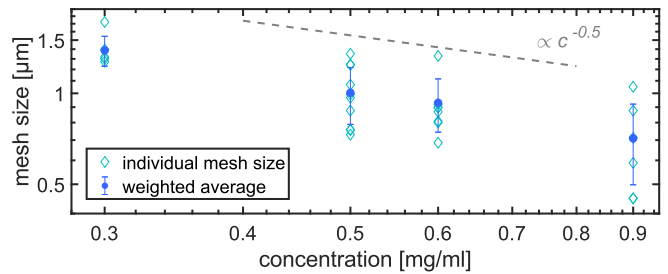}
  \caption{Mesh size vs. monomer concentration for F-actin networks crosslinked with wLX. Error bars are calculated from relative error propagation and weighted averaging over several individual tracers \cite{supplemental}. The expected scaling $\xi \propto c^{-0.5}$ \cite{de_gennes_remarks_1976, isambert_dynamics_1996} is indicated by the dashed line.}
  \label{fig_fig3}
\end{figure}

Both experimental and theoretical polymer physics study how features of the constituting elements determine the properties of whole polymer networks. Previous studies have examined the interaction of semiflexible filaments \cite{kas_direct_1994} or investigated how geometrical parameters \cite{schuldt_tuning_2016, keshavarz_confining_2017} and bulk mechanical properties \cite{keshavarz_nanoscale_2016} can be deduced from tracking single filaments embedded in an entangled network.
However, complex semiflexible polymer networks such as the cellular cytoskeleton contain multiple types of filaments with varied mechanical properties as well as crosslinkers which create physical connections between filaments \cite{huber_emergent_2013}. A systematic investigation of the impact from these additional elements upon the dynamics of a single filament moving within the background network has not been possible with the systems typically used for reptation measurements. In our study, we use semiflexible DNA nanotubes with tunable stiffness \cite{yin_programming_2008} as embedded tracers in crosslinked and entangled F-actin networks and DNA nanotube networks \cite{schuldt_tuning_2016}. Since the DNA filaments do not specifically interact with the network constituents, we are able to observe single-filament dynamics and directly measure the tube width of both entangled and crosslinked F-actin, as well as DNA nanotube networks, with respect to tracer stiffness.
By applying principles of the tube model \cite{hinsch_quantitative_2007}, we determined the mesh size of these polymer networks and confirmed some basic theoretical predictions such as the scaling of mesh size with monomer concentration. Although the concept has been proven to be successful in deriving scaling laws that are experimentally accessible \cite{mackintosh_elasticity_1995, isambert_dynamics_1996, broedersz_modeling_2014}, we find a need for adjustment beyond solely accounting for steric interactions, since the predictions dependent on tracer stiffness are not supported by our data. Nonetheless, our approach allowed us to study the architecture of crosslinked networks, opening new venues to investigate a broad range of filamentous networks by applying tube model predictions.\\
In this framework, the difficulties in describing semiflexible polymer networks as a many-body problem are reduced by studying a single test filament in the background of all other filaments [see Fig. \ref{fig_fig1}(a)]. These background filaments constrain the test filament so that its effective available space is realized by a curved cylindrical tube \cite{edwards_statistical_1967}, wherein the filament can reptate, i.e., diffuse along its own contour \cite{gennes_reptation_1971}. The mesh size $\xi$ of a semiflexible polymer network is defined as the average distance between two filaments and, thus, depends on the monomer concentration of the solution, $\xi \propto 1/c^2$ \cite{de_gennes_remarks_1976, isambert_dynamics_1996}. We studied networks usually termed as concentrated solutions or tightly-entangled polymer networks, where the diameter \(a\) of the reptation tube is assumed to be much smaller than the persistence length \(l_{\text{p}}\) of the constituting filaments, $a \lesssim \xi \ll l_{\text{p}} \approx L$ \cite{odijk_statistics_1983, semenov_dynamics_1986,isambert_dynamics_1996, morse_tube_2001,hinsch_quantitative_2007,tassieri_dynamics_2017}, with $L$ denoting the contour length of the tracers [see Fig. \ref{fig_fig1}(a)].
Considering the relationship of tube width $a$, mesh size $\xi$, and persistence length $l_{\text{p}}$, the same scaling law has been obtained by different argumentation, \(a \propto \frac{\xi^{6/5}}{l_{\text{p}}^{1/5}}\) \cite{semenov_dynamics_1986,morse_tube_2001,hinsch_quantitative_2007}.\\
This relation alone is insufficient to measure a network's mesh size quantitatively by detecting the tube width of embedded tracer filaments. \citet{hinsch_quantitative_2007} calculated a prefactor of 0.31 in a self-consistent treatment of the network by allowing fluctuations of the background filaments and decomposing the test filament into independent rods of appropriate length. They also found an additional term referring to boundary effects at the tube ends, using an expression for the free energy of a confined semiflexible polymer of finite length \cite{burkhardt_free_1995}. This second-order term accounts for the fact that short tracer filaments are more influenced by finite-length effects than long tracers. The complete expression reads \cite{hinsch_quantitative_2007}
\begin{equation}
a \approx 0.31\frac{\xi^{6/5}}{l_{\text{p}}^{1/5}} + 0.59\frac{\xi^2}{L}.
\label{eq_xi_hinsch}
\end{equation}
Equation \eqref{eq_xi_hinsch} was originally derived for entangled polymer networks, but we were also able to employ it for the determination of the mesh size of crosslinked F-actin networks by using DNA nanotubes as embedded tracer filaments. These filaments have no binding sites for explicit actin crosslinking complexes and are thus usable as decoupled, reptating tracers in crosslinked F-actin networks. DNA nanotubes have the additional benefit that their persistence length can be varied sufficiently to investigate theoretical predictions such as $a \propto l_{\text{p}}^{-0.2}$, which were not experimentally accessible before \cite{schuldt_tuning_2016}.\\
DNA nanotubes were hybridized from $n$ partially complementary oligonucleotides, forming \(n\)-helix tubes (\(n\)HTs) [see Fig. \ref{fig_fig1}(b)]. Depending on the chosen set of \(n\) oligonucleotides, \(n\)HTs with different diameters and, accordingly, persistence lengths are formed \cite{yin_programming_2008, schiffels_nanoscale_2013, schuldt_tuning_2016, glaser_self_2016}. Their contour length distribution is comparable to that of actin filaments, while the filament width stays below \SI{10}{\nano\meter} \cite{yin_programming_2008}. To hybridize fluorescent tracer filaments, a fluorescent dye was attached to one of the oligonucleotides [see Fig. \ref{fig_fig1}(c)]. After hybridization, \(n\)HTs are stable for weeks and maintain their structure if they are kept below the structure's melting temperature of approximately \SI{60}{\celsius} \cite{yin_programming_2008}. Subsequently, we were able to polymerize F-actin background networks around the tracer filaments, ensuring them to be homogeneously distributed.\\
For the crosslinked F-actin background network, we selected the synthetic crosslinker wLX (weak LifeAct\textsuperscript{\textregistered}-based crosslinker) introduced by \citet{lorenz_synthetic_2018}. It consists of two actin-binding LifeAct\textsuperscript{\textregistered} peptides connected by double-stranded DNA and transiently links actin filaments, forming networks that mechanically resemble networks of F-actin and the natural crosslinker \(\alpha\)-actinin \cite{lorenz_synthetic_2018}. We preferred using wLX over naturally occurring crosslinkers because of its reproducibility, i.e., defined length and binding strength, and applicability, as the fluorescent \(n\)HT tracers showed unspecific interactions with natural crosslinkers such as \(\alpha\)-actinin and heavy meromyosin (data not shown) that could be avoided by using wLX. The ratio of actin monomers to crosslinker molecules was chosen carefully to 150:1 so that the network was fully crosslinked, but not bundled \cite{lorenz_synthetic_2018}.\\
In order to compare the reptation of \(n\)HTs in F-actin networks to the behavior of \(n\)HTs in nanotube networks, we chose 8HT as the constituting background filament type since its persistence length of \SI{8.9}{\micro\meter} is comparable to that of actin filaments \cite{isambert_flexibility_1995}. Likewise, we used 8HT as tracer filaments in F-actin background networks of similar persistence length.\\
The reptation of \(n\)HT tracers in different background networks was observed using epi-fluorescence microscopy and tracked with the ImageJ plugin JFilament \cite{smith_segmentation_2010}, see Fig. \ref{fig_fig1}(d). Details on the hybridization process, sample preparation, measurements, and data processing are given in the Supplemental Material \cite{supplemental}. By analyzing multiple subsets of 100 frames, we were able to determine a tube width for each individual tracer filament.\\
After calculating a weighted mean for all tube widths measured with one type of DNA nanotube tracer, we carried out an error-weighted fit to determine the power law exponents of the tube width scaling with the tracers' persistence length. Figure \ref{fig_fig2} shows the measured tube width values for five types of \(n\)HT filaments with persistence lengths ranging from approximately \SI{3}{\micro\meter} to \SI{13}{\micro\meter} in three distinct background networks.\\
In general, one would expect to observe different tube widths for entangled and for crosslinked background networks of the same concentration since some fluctuation modes of the background polymers are suppressed in crosslinked networks \cite{marrucci_relaxation_1985}. However, we observe nearly the same values for tube widths in entangled and crosslinked F-actin networks, indicating that the crosslinker wLX does not change the network geometry for the chosen ratio of actin monomers to crosslinker molecules \cite{lorenz_synthetic_2018}. In 8HT networks, \(n\)HTs were less motile than in F-actin networks and appeared nearly stuck, resulting in lower tube widths.\\
The predicted scaling $a \propto l_{\text{p}}^{-0.2}$ (indicated by the dashed line in Fig. \ref{fig_fig2}) could not be confirmed in all of the three background networks. It is only observed for \(n\)HTs in entangled F-actin where the exponent is \SI{-0.33\pm 0.16}{} so that an exponent of \(-0.2\) is within the error limits. The power law exponent for crosslinked F-actin is \SI{-0.42\pm 0.08}{}. However, since the tube width values have relatively large errors, it cannot be concluded that the predicted scaling does not apply in this case. The scaling exponent for 8HT as a background network is \SI{-0.74\pm 0.12}{} which differs from the predicted exponent by a factor of at least 3. Together with the observed lower motility of \(n\)HTs in 8HT background, this implies unspecific interactions between the filaments that are not captured by the tube model.\\
By fitting the leading term of Eq. \eqref{eq_xi_hinsch}, we were also able to derive the mesh size of the background networks. The estimated mesh sizes are \SI{1}{\micro\meter} for entangled F-actin, \SI{1.2}{\micro\meter} for crosslinked F-actin and \SI{1.4}{\micro\meter} for 8HT networks, corresponding well with previous publications \cite{schuldt_tuning_2016, golde_glassy_2018}.\\
With our DNA nanotube tracers, we were able to measure the mesh size of crosslinked F-actin networks. This is not possible using actin-based tracer filaments because they do not reptate in the presence of actin crosslinkers. 8HTs were again chosen for their persistence length comparability to unstabilized actin filaments \cite{isambert_flexibility_1995}, avoiding effects stemming from different polymer stiffnesses. By measuring the tube width of embedded 8HTs and employing Eq. \eqref{eq_xi_hinsch} with the adjusted scaling of $a \propto l_{\text{p}}^{-0.42}$ obtained from the fit plotted in Fig. \ref{fig_fig2}, we determined the mesh size for crosslinked F-actin networks of four different concentrations (see Fig. \ref{fig_fig3}). The expected scaling $\xi \propto c^{-0.5}$ \cite{de_gennes_remarks_1976, isambert_dynamics_1996} was confirmed as it has already been for entangled F-actin \cite{schmidt_chain_1989, schuldt_tuning_2016}.
The resulting mesh size of \SI{1\pm 0.2}{\micro\meter} for a concentration of \SI{0.5}{\milli\gram\per\milli\litre} is comparable to the fitted mesh size of crosslinked F-actin networks from Fig. \ref{fig_fig2} and to already published values for entangled F-actin of the same concentration \cite{schuldt_tuning_2016, golde_glassy_2018}, implying that crosslinking does not change the geometry of the F-actin network.\\
Since filaments with a shorter contour length $L$ are more affected by boundary effects at the end of the tube, we examined whether the scaling of the second-order term $a \propto L^{-1}$ of Eq. \eqref{eq_xi_hinsch} applies to our data.
\begin{figure}
        \includegraphics[width=\columnwidth]{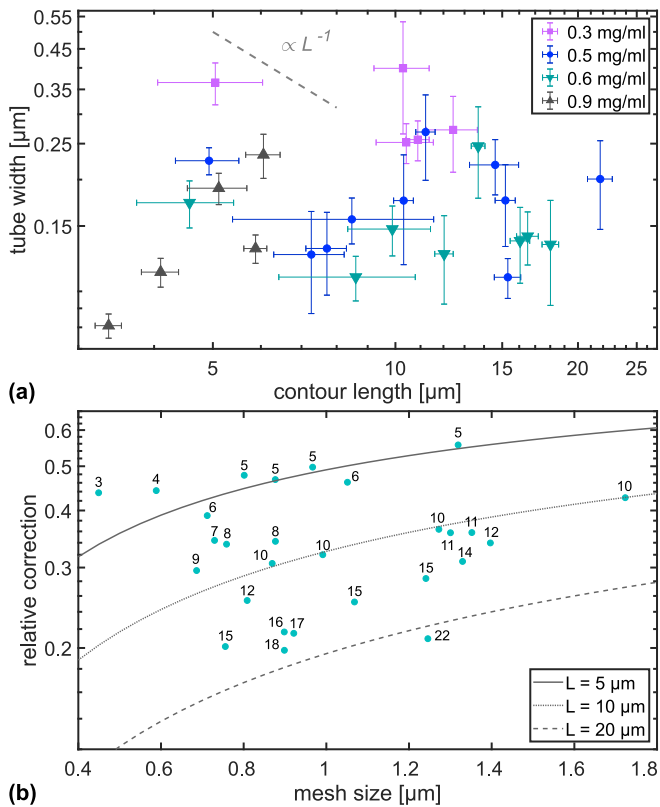}
         \caption{(a) Individual tube widths for 8HTs in crosslinked F-actin background networks versus tracer contour lengths. The color coding refers to the networks' actin monomer concentration. The predicted scaling $a \propto L^{-1}$  is indicated by the dashed line. Error bars are standard errors \cite{supplemental}. (b) This plot shows the relative correction $(0.59\frac{\xi^2}{L})\frac{1}{a}$ versus mesh size (circles, calculated by $a \propto 0.31\frac{\xi^{6/5}}{l_{\text{p}}^{0.42}} + 0.59\frac{\xi^2}{L}$) for the data of panel (a). The numbers accompanying each data point are the filaments' rounded contour lengths. Predictive curves of relative corrections are plotted for theoretical contour lengths of \SI{5}{\micro\meter}, \SI{10}{\micro\meter} and \SI{20}{\micro\meter}.}
         \label{fig_fig4}
\end{figure}
We chose to study this aspect by measuring tube widths and contour lengths of embedded 8HTs in crosslinked F-actin background networks at various concentrations. In Fig. \ref{fig_fig4}(a), all resulting individual tube widths for 8HTs are plotted over the tracers' contour lengths. The predicted scaling $a \propto L^{-1}$ (indicated by the dashed line) is not fulfilled for any of the background network concentrations. However, the fact that we do not observe it does not mean that the filaments are not affected by the predicted boundary effects. The leading term of Eq. \eqref{eq_xi_hinsch}, by itself, is valid for infinite tracer filaments; the second term originates from considering the free energy of finite tracer filaments constrained by the surrounding network and corrects for polydisperse tracers \cite{hinsch_quantitative_2007}.\\
To quantify the effect of such finite tracer lengths, we computed the relative correction of the tube width stemming from the second term of Eq. \eqref{eq_xi_hinsch}. The results plotted in Fig. \ref{fig_fig4}(b) reveal that the relative correction is indeed higher for shorter filaments. Figure \ref{fig_fig4}(b) also includes calculated curves of relative corrections for distinct contour lengths analogous to predictions from \citet{hinsch_quantitative_2007}. The data points match these predictions well, illustrating that longer filaments are less affected by boundary affects at the end of the tube. However, the observed effect of these relative corrections of tube widths is not reflected in the variation of the absolute values of tube widths, probably because local inhomogeneities affecting each individual tracer's observed tube width have a higher influence than the finite-length effect.
Figure \ref{fig_fig4} shows the tube widths and corrections for crosslinked F-actin networks under the condition that the tracer persistence length equals that of the background network filaments. Repeating this analysis lead to the same conclusion for all other sets of tube width data from \(n\)HT tracers embedded in three different background networks \cite{supplemental}.\\
Although we expected to see a difference between the tube widths of entangled and crosslinked F-actin networks at the same concentration due to the suppression of fluctuation modes of crosslinked background filaments, the average values were comparable within the error bars, indicating that the fluctuations of background filaments do not have a high impact on the constraining potential exerted on the tracer filament \cite{wang_confining_2010, keshavarz_confining_2017}. By varying the persistence length $l_{\text{p}}$ of the tracers, we found that the expected scaling relation $a \propto l_{\text{p}}^{-0.2}$ is not valid for all of the three examined background networks. 8HT background networks show a different relation between tube width and persistence length, presumably due to interactions between the \(n\)HT tracers and the 8HT network that are not accounted for in the general reptation model, e.g., electrostatic interactions between the DNA molecules that lead to an effective stickiness \cite{cherstvy_electrostatic_2011}. Recent studies have shown that other scaling predictions made from the tube model are not applicable to networks of \(n\)HTs \cite{schuldt_tuning_2016} and that sticky interactions lead to altered bulk properties of semiflexible polymer networks \cite{golde_role_2019}. We speculate that an experimental adjustment of the exponent in the relation $a \propto l_{\text{p}}^{-0.2}$ is acceptable when the interactions not included in the tube model only mildly affect the tracers' reptation.\\
Previous publications examining reptation have made use of F-actin tracers in entangled F-actin \cite{kas_direct_1994, glaser_tube_2010}. Stiff filaments \cite{fakhri_brownian_2010} and flexible tracers \cite{keshavarz_nanoscale_2016, keshavarz_confining_2017} have also been measured. There is no experimental study that tests the predicted scaling $a \propto l_{\text{p}}^{-0.2}$ for semiflexible filaments over a broad range of $l_{\text{p}}$. A recent attempt by \citet{keshavarz_confining_2017} utilized flexible filaments with different persistence lengths, but was limited to two data points, hindering a clear verification of a scaling law. However, the same study proved the prediction \(a \propto \xi^{6/5}\) for flexible filaments where \(\xi\) can be calculated independently of \(a\) from the elastic plateau modulus obtained by rheological measurements \cite{keshavarz_confining_2017}. In turn, we were able to measure the mesh size of crosslinked F-actin networks by applying Eq. \eqref{eq_xi_hinsch}. Until now, this had only been done before for entangled F-actin by employing actin filaments as tracers \cite{schuldt_tuning_2016, golde_glassy_2018}. The results for crosslinked F-actin networks reveal a good agreement with the established concentration scaling $\xi \propto c^{-0.5}$ \cite{schmidt_chain_1989, schuldt_tuning_2016} and the absolute values are well comparable to those published for entangled F-actin networks \cite{schuldt_tuning_2016, golde_glassy_2018}, confirming that the crosslinker wLX does not change the network architecture for the chosen crosslinker concentration \cite{lorenz_synthetic_2018}.\\
The correspondence of measured mesh size values between entangled and crosslinked F-actin networks justifies the use of Eq. \eqref{eq_xi_hinsch} for the crosslinked networks, even if it was originally derived for an entangled background network with fluctuating background filaments. This is comparable to the findings of \citet{gardel_elastic_2004}, where the elastic modulus of a crosslinked F-actin network followed the concentration scaling that had been deduced for an entangled network in the tube model framework previously \cite{mackintosh_elasticity_1995}. We would like to point out that the use of other actin crosslinkers might change the network architecture or the interactions between background network and tracer filaments in such a way that it is not possible to use Eq. \eqref{eq_xi_hinsch} for data evaluation.\\
A recent simulation study investigated the difference between static and entangled background polymers and its effect on stress relaxation in the framework of the tube model \cite{lang_disentangling_2018}. \citet{lang_disentangling_2018} showed that constraint release, the dissolution of the reptation tube upon correlated motion of the tracer and its surrounding filaments, leads to altered tracer dynamics and relaxation times \cite{isambert_dynamics_1996, keshavarz_nanoscale_2016, lang_disentangling_2018}. This could explain why the conventional tube model predictions do not fully account for the examined networks, which differ from a structure with fixed obstacles.\\
In conclusion, we have tested the persistence length dependency of tracer filaments reptating in a background network, revealing that the tube model needs to be extended to describe actual experimental data, where more than steric interactions are present. However, the tube model does indeed provide valid predictions for measuring parameters describing the geometrical structure. Utilizing mechanically tunable filaments as tracers, not only entangled, but also crosslinked networks can be architecturally characterized. This approach is particularly appealing because it avoids additional influences inherent to embedding fluorescent beads for these determinations \cite{schmidt_chain_1989, golde_fluorescent_2013} or labeling the entire network with fluorophores \cite{fischer_fast_2019}. Our method may also prove applicable to other semiflexible polymer networks, e.g., networks of other cytoskeletal filaments or collagen fibers that have been physically crosslinked. The programmability of DNA nanotubes further allows one to choose a suitable tracer’s persistence length to match that of the background network of interest, thus mitigating emergent effects due to different polymer stiffnesses. Furthermore, stable and biocompatible probe filaments like DNA nanotubes may be utilized to measure local properties of heterogeneous scaffolds such as the extracellular matrix or the cellular cytoskeleton.\\
\begin{acknowledgments}
We thank Josef A. Käs for fruitful discussions.\\
We acknowledge funding by the European Research Council (ERC-741350). C.T. acknowledges funding from the European Social Fund (ESF—100380880).
\end{acknowledgments}

T.H. and C.T. contributed equally to this work.


\begin{thebibliography}{35}%
\makeatletter
\providecommand \@ifxundefined [1]{%
 \@ifx{#1\undefined}
}%
\providecommand \@ifnum [1]{%
 \ifnum #1\expandafter \@firstoftwo
 \else \expandafter \@secondoftwo
 \fi
}%
\providecommand \@ifx [1]{%
 \ifx #1\expandafter \@firstoftwo
 \else \expandafter \@secondoftwo
 \fi
}%
\providecommand \natexlab [1]{#1}%
\providecommand \enquote  [1]{``#1''}%
\providecommand \bibnamefont  [1]{#1}%
\providecommand \bibfnamefont [1]{#1}%
\providecommand \citenamefont [1]{#1}%
\providecommand \href@noop [0]{\@secondoftwo}%
\providecommand \href [0]{\begingroup \@sanitize@url \@href}%
\providecommand \@href[1]{\@@startlink{#1}\@@href}%
\providecommand \@@href[1]{\endgroup#1\@@endlink}%
\providecommand \@sanitize@url [0]{\catcode `\\12\catcode `\$12\catcode
  `\&12\catcode `\#12\catcode `\^12\catcode `\_12\catcode `\%12\relax}%
\providecommand \@@startlink[1]{}%
\providecommand \@@endlink[0]{}%
\providecommand \url  [0]{\begingroup\@sanitize@url \@url }%
\providecommand \@url [1]{\endgroup\@href {#1}{\urlprefix }}%
\providecommand \urlprefix  [0]{URL }%
\providecommand \Eprint [0]{\href }%
\providecommand \doibase [0]{http://dx.doi.org/}%
\providecommand \selectlanguage [0]{\@gobble}%
\providecommand \bibinfo  [0]{\@secondoftwo}%
\providecommand \bibfield  [0]{\@secondoftwo}%
\providecommand \translation [1]{[#1]}%
\providecommand \BibitemOpen [0]{}%
\providecommand \bibitemStop [0]{}%
\providecommand \bibitemNoStop [0]{.\EOS\space}%
\providecommand \EOS [0]{\spacefactor3000\relax}%
\providecommand \BibitemShut  [1]{\csname bibitem#1\endcsname}%
\let\auto@bib@innerbib\@empty
\bibitem [{\citenamefont {Käs}\ \emph {et~al.}(1994)\citenamefont {Käs},
  \citenamefont {Strey},\ and\ \citenamefont {Sackmann}}]{kas_direct_1994}%
  \BibitemOpen
  \bibfield  {author} {\bibinfo {author} {\bibfnamefont {J.}~\bibnamefont
  {Käs}}, \bibinfo {author} {\bibfnamefont {H.}~\bibnamefont {Strey}}, \ and\
  \bibinfo {author} {\bibfnamefont {E.}~\bibnamefont {Sackmann}},\ }\bibfield
  {title} {\enquote {\bibinfo {title} {Direct imaging of reptation for
  semiflexible actin filaments},}\ }\href {\doibase 10.1038/368226a0}
  {\bibfield  {journal} {\bibinfo  {journal} {Nature}\ }\textbf {\bibinfo
  {volume} {368}},\ \bibinfo {pages} {226--229} (\bibinfo {year}
  {1994})}\BibitemShut {NoStop}%
\bibitem [{\citenamefont {Schuldt}\ \emph {et~al.}(2016)\citenamefont
  {Schuldt}, \citenamefont {Schnau\ss}, \citenamefont {H\"andler},
  \citenamefont {Glaser}, \citenamefont {Lorenz}, \citenamefont {Golde},
  \citenamefont {K\"as},\ and\ \citenamefont {Smith}}]{schuldt_tuning_2016}%
  \BibitemOpen
  \bibfield  {author} {\bibinfo {author} {\bibfnamefont {C.}~\bibnamefont
  {Schuldt}}, \bibinfo {author} {\bibfnamefont {J.}~\bibnamefont {Schnau\ss}},
  \bibinfo {author} {\bibfnamefont {T.}~\bibnamefont {H\"andler}}, \bibinfo
  {author} {\bibfnamefont {M.}~\bibnamefont {Glaser}}, \bibinfo {author}
  {\bibfnamefont {J.}~\bibnamefont {Lorenz}}, \bibinfo {author} {\bibfnamefont
  {T.}~\bibnamefont {Golde}}, \bibinfo {author} {\bibfnamefont {J.~A.}\
  \bibnamefont {K\"as}}, \ and\ \bibinfo {author} {\bibfnamefont {D.~M.}\
  \bibnamefont {Smith}},\ }\bibfield  {title} {\enquote {\bibinfo {title}
  {Tuning {Synthetic} {Semiflexible} {Networks} by {Bending} {Stiffness}},}\
  }\href {\doibase 10.1103/PhysRevLett.117.197801} {\bibfield  {journal}
  {\bibinfo  {journal} {Physical Review Letters}\ }\textbf {\bibinfo {volume}
  {117}},\ \bibinfo {pages} {197801} (\bibinfo {year} {2016})}\BibitemShut
  {NoStop}%
\bibitem [{\citenamefont {Keshavarz}\ \emph {et~al.}(2017)\citenamefont
  {Keshavarz}, \citenamefont {Engelkamp}, \citenamefont {Xu}, \citenamefont
  {van~den Boomen}, \citenamefont {Maan}, \citenamefont {Christianen},\ and\
  \citenamefont {Rowan}}]{keshavarz_confining_2017}%
  \BibitemOpen
  \bibfield  {author} {\bibinfo {author} {\bibfnamefont {M.}~\bibnamefont
  {Keshavarz}}, \bibinfo {author} {\bibfnamefont {H.}~\bibnamefont
  {Engelkamp}}, \bibinfo {author} {\bibfnamefont {J.}~\bibnamefont {Xu}},
  \bibinfo {author} {\bibfnamefont {O.~I.}\ \bibnamefont {van~den Boomen}},
  \bibinfo {author} {\bibfnamefont {J.~C.}\ \bibnamefont {Maan}}, \bibinfo
  {author} {\bibfnamefont {P.~C.~M.}\ \bibnamefont {Christianen}}, \ and\
  \bibinfo {author} {\bibfnamefont {A.~E.}\ \bibnamefont {Rowan}},\ }\bibfield
  {title} {\enquote {\bibinfo {title} {Confining {Potential} as a {Function} of
  {Polymer} {Stiffness} and {Concentration} in {Entangled} {Polymer}
  {Solutions}},}\ }\href {\doibase 10.1021/acs.jpcb.6b12667} {\bibfield
  {journal} {\bibinfo  {journal} {The Journal of Physical Chemistry B}\
  }\textbf {\bibinfo {volume} {121}},\ \bibinfo {pages} {5613--5620} (\bibinfo
  {year} {2017})}\BibitemShut {NoStop}%
\bibitem [{\citenamefont {Keshavarz}\ \emph {et~al.}(2016)\citenamefont
  {Keshavarz}, \citenamefont {Engelkamp}, \citenamefont {Xu}, \citenamefont
  {Braeken}, \citenamefont {Otten}, \citenamefont {Uji-i}, \citenamefont
  {Schwartz}, \citenamefont {Koepf}, \citenamefont {Vananroye}, \citenamefont
  {Vermant}, \citenamefont {Nolte}, \citenamefont {De~Schryver}, \citenamefont
  {Maan}, \citenamefont {Hofkens}, \citenamefont {Christianen},\ and\
  \citenamefont {Rowan}}]{keshavarz_nanoscale_2016}%
  \BibitemOpen
  \bibfield  {author} {\bibinfo {author} {\bibfnamefont {M.}~\bibnamefont
  {Keshavarz}}, \bibinfo {author} {\bibfnamefont {H.}~\bibnamefont
  {Engelkamp}}, \bibinfo {author} {\bibfnamefont {J.}~\bibnamefont {Xu}},
  \bibinfo {author} {\bibfnamefont {E.}~\bibnamefont {Braeken}}, \bibinfo
  {author} {\bibfnamefont {M.~B.~J.}\ \bibnamefont {Otten}}, \bibinfo {author}
  {\bibfnamefont {H.}~\bibnamefont {Uji-i}}, \bibinfo {author} {\bibfnamefont
  {E.}~\bibnamefont {Schwartz}}, \bibinfo {author} {\bibfnamefont
  {M.}~\bibnamefont {Koepf}}, \bibinfo {author} {\bibfnamefont
  {A.}~\bibnamefont {Vananroye}}, \bibinfo {author} {\bibfnamefont
  {J.}~\bibnamefont {Vermant}}, \bibinfo {author} {\bibfnamefont {R.~J.~M.}\
  \bibnamefont {Nolte}}, \bibinfo {author} {\bibfnamefont {F.}~\bibnamefont
  {De~Schryver}}, \bibinfo {author} {\bibfnamefont {J.~C.}\ \bibnamefont
  {Maan}}, \bibinfo {author} {\bibfnamefont {J.}~\bibnamefont {Hofkens}},
  \bibinfo {author} {\bibfnamefont {P.~C.~M.}\ \bibnamefont {Christianen}}, \
  and\ \bibinfo {author} {\bibfnamefont {A.~E.}\ \bibnamefont {Rowan}},\
  }\bibfield  {title} {\enquote {\bibinfo {title} {Nanoscale {Study} of
  {Polymer} {Dynamics}},}\ }\href {\doibase 10.1021/acsnano.5b06931} {\bibfield
   {journal} {\bibinfo  {journal} {ACS Nano}\ }\textbf {\bibinfo {volume}
  {10}},\ \bibinfo {pages} {1434--1441} (\bibinfo {year} {2016})}\BibitemShut
  {NoStop}%
\bibitem [{\citenamefont {Huber}\ \emph {et~al.}(2013)\citenamefont {Huber},
  \citenamefont {Schnau\ss}, \citenamefont {Rönicke}, \citenamefont {Rauch},
  \citenamefont {Müller}, \citenamefont {F\"utterer},\ and\ \citenamefont
  {K\"as}}]{huber_emergent_2013}%
  \BibitemOpen
  \bibfield  {author} {\bibinfo {author} {\bibfnamefont {F.}~\bibnamefont
  {Huber}}, \bibinfo {author} {\bibfnamefont {J.}~\bibnamefont {Schnau\ss}},
  \bibinfo {author} {\bibfnamefont {S.}~\bibnamefont {Rönicke}}, \bibinfo
  {author} {\bibfnamefont {P.}~\bibnamefont {Rauch}}, \bibinfo {author}
  {\bibfnamefont {K.}~\bibnamefont {Müller}}, \bibinfo {author} {\bibfnamefont
  {C.}~\bibnamefont {F\"utterer}}, \ and\ \bibinfo {author} {\bibfnamefont
  {J.~A.}\ \bibnamefont {K\"as}},\ }\bibfield  {title} {\enquote {\bibinfo
  {title} {Emergent complexity of the cytoskeleton: from single filaments to
  tissue},}\ }\href {https://doi.org/10.1080/00018732.2013.771509} {\bibfield
  {journal} {\bibinfo  {journal} {Advances in Physics}\ }\textbf {\bibinfo
  {volume} {62}},\ \bibinfo {pages} {1--112} (\bibinfo {year}
  {2013})}\BibitemShut {NoStop}%
\bibitem [{\citenamefont {Yin}\ \emph {et~al.}(2008)\citenamefont {Yin},
  \citenamefont {Hariadi}, \citenamefont {Sahu}, \citenamefont {Choi},
  \citenamefont {Park}, \citenamefont {LaBean},\ and\ \citenamefont
  {Reif}}]{yin_programming_2008}%
  \BibitemOpen
  \bibfield  {author} {\bibinfo {author} {\bibfnamefont {P.}~\bibnamefont
  {Yin}}, \bibinfo {author} {\bibfnamefont {R.~F.}\ \bibnamefont {Hariadi}},
  \bibinfo {author} {\bibfnamefont {S.}~\bibnamefont {Sahu}}, \bibinfo {author}
  {\bibfnamefont {H.~M.~T.}\ \bibnamefont {Choi}}, \bibinfo {author}
  {\bibfnamefont {S.~H.}\ \bibnamefont {Park}}, \bibinfo {author}
  {\bibfnamefont {T.~H.}\ \bibnamefont {LaBean}}, \ and\ \bibinfo {author}
  {\bibfnamefont {J.~H.}\ \bibnamefont {Reif}},\ }\bibfield  {title} {\enquote
  {\bibinfo {title} {Programming {DNA} {Tube} {Circumferences}},}\ }\href
  {\doibase 10.1126/science.1157312} {\bibfield  {journal} {\bibinfo  {journal}
  {Science}\ }\textbf {\bibinfo {volume} {321}},\ \bibinfo {pages} {824--826}
  (\bibinfo {year} {2008})}\BibitemShut {NoStop}%
\bibitem [{\citenamefont {Hinsch}\ \emph {et~al.}(2007)\citenamefont {Hinsch},
  \citenamefont {Wilhelm},\ and\ \citenamefont
  {Frey}}]{hinsch_quantitative_2007}%
  \BibitemOpen
  \bibfield  {author} {\bibinfo {author} {\bibfnamefont {H.}~\bibnamefont
  {Hinsch}}, \bibinfo {author} {\bibfnamefont {J.}~\bibnamefont {Wilhelm}}, \
  and\ \bibinfo {author} {\bibfnamefont {E.}~\bibnamefont {Frey}},\ }\bibfield
  {title} {\enquote {\bibinfo {title} {Quantitative tube model for semiflexible
  polymer solutions},}\ }\href {\doibase 10.1140/epje/i2007-10208-2} {\bibfield
   {journal} {\bibinfo  {journal} {The European Physical Journal E}\ }\textbf
  {\bibinfo {volume} {24}},\ \bibinfo {pages} {35--46} (\bibinfo {year}
  {2007})}\BibitemShut {NoStop}%
\bibitem [{\citenamefont {MacKintosh}\ \emph {et~al.}(1995)\citenamefont
  {MacKintosh}, \citenamefont {K\"as},\ and\ \citenamefont
  {Janmey}}]{mackintosh_elasticity_1995}%
  \BibitemOpen
  \bibfield  {author} {\bibinfo {author} {\bibfnamefont {F.~C.}\ \bibnamefont
  {MacKintosh}}, \bibinfo {author} {\bibfnamefont {J.}~\bibnamefont {K\"as}}, \
  and\ \bibinfo {author} {\bibfnamefont {P.~A.}\ \bibnamefont {Janmey}},\
  }\bibfield  {title} {\enquote {\bibinfo {title} {Elasticity of semiflexible
  biopolymer networks},}\ }\href {\doibase 10.1103/PhysRevLett.75.4425}
  {\bibfield  {journal} {\bibinfo  {journal} {Phys. Rev. Lett.}\ }\textbf
  {\bibinfo {volume} {75}},\ \bibinfo {pages} {4425--4428} (\bibinfo {year}
  {1995})}\BibitemShut {NoStop}%
\bibitem [{\citenamefont {Isambert}\ and\ \citenamefont
  {Maggs}(1996)}]{isambert_dynamics_1996}%
  \BibitemOpen
  \bibfield  {author} {\bibinfo {author} {\bibfnamefont {H.}~\bibnamefont
  {Isambert}}\ and\ \bibinfo {author} {\bibfnamefont {A.~C.}\ \bibnamefont
  {Maggs}},\ }\bibfield  {title} {\enquote {\bibinfo {title} {Dynamics and
  {Rheology} of {Actin} {Solutions}},}\ }\href {\doibase 10.1021/ma946418x}
  {\bibfield  {journal} {\bibinfo  {journal} {Macromolecules}\ }\textbf
  {\bibinfo {volume} {29}},\ \bibinfo {pages} {1036--1040} (\bibinfo {year}
  {1996})}\BibitemShut {NoStop}%
\bibitem [{\citenamefont {Broedersz}\ and\ \citenamefont
  {MacKintosh}(2014)}]{broedersz_modeling_2014}%
  \BibitemOpen
  \bibfield  {author} {\bibinfo {author} {\bibfnamefont {C.~P.}\ \bibnamefont
  {Broedersz}}\ and\ \bibinfo {author} {\bibfnamefont {F.~C.}\ \bibnamefont
  {MacKintosh}},\ }\bibfield  {title} {\enquote {\bibinfo {title} {Modeling
  semiflexible polymer networks},}\ }\href {\doibase 10.1103/RevModPhys.86.995}
  {\bibfield  {journal} {\bibinfo  {journal} {Rev. Mod. Phys.}\ }\textbf
  {\bibinfo {volume} {86}},\ \bibinfo {pages} {995--1036} (\bibinfo {year}
  {2014})}\BibitemShut {NoStop}%
\bibitem [{\citenamefont {Edwards}(1967)}]{edwards_statistical_1967}%
  \BibitemOpen
  \bibfield  {author} {\bibinfo {author} {\bibfnamefont {S.~F.}\ \bibnamefont
  {Edwards}},\ }\bibfield  {title} {\enquote {\bibinfo {title} {The statistical
  mechanics of polymerized material},}\ }\href {\doibase
  10.1088/0370-1328/92/1/303} {\bibfield  {journal} {\bibinfo  {journal}
  {Proceedings of the Physical Society}\ }\textbf {\bibinfo {volume} {92}},\
  \bibinfo {pages} {9--16} (\bibinfo {year} {1967})}\BibitemShut {NoStop}%
\bibitem [{\citenamefont {de~Gennes}(1971)}]{gennes_reptation_1971}%
  \BibitemOpen
  \bibfield  {author} {\bibinfo {author} {\bibfnamefont {P.~G.}\ \bibnamefont
  {de~Gennes}},\ }\bibfield  {title} {\enquote {\bibinfo {title} {Reptation of
  a {Polymer} {Chain} in the {Presence} of {Fixed} {Obstacles}},}\ }\href
  {\doibase 10.1063/1.1675789} {\bibfield  {journal} {\bibinfo  {journal} {The
  Journal of Chemical Physics}\ }\textbf {\bibinfo {volume} {55}},\ \bibinfo
  {pages} {572--579} (\bibinfo {year} {1971})}\BibitemShut {NoStop}%
\bibitem [{\citenamefont {de~Gennes}\ \emph {et~al.}(1976)\citenamefont
  {de~Gennes}, \citenamefont {Pincus}, \citenamefont {Velasco},\ and\
  \citenamefont {Brochard}}]{de_gennes_remarks_1976}%
  \BibitemOpen
  \bibfield  {author} {\bibinfo {author} {\bibfnamefont {P.~G.}\ \bibnamefont
  {de~Gennes}}, \bibinfo {author} {\bibfnamefont {P.}~\bibnamefont {Pincus}},
  \bibinfo {author} {\bibfnamefont {R.~M.}\ \bibnamefont {Velasco}}, \ and\
  \bibinfo {author} {\bibfnamefont {F.}~\bibnamefont {Brochard}},\ }\bibfield
  {title} {\enquote {\bibinfo {title} {Remarks on polyelectrolyte
  conformation},}\ }\href {\doibase 10.1051/jphys:0197600370120146100}
  {\bibfield  {journal} {\bibinfo  {journal} {Journal de Physique}\ }\textbf
  {\bibinfo {volume} {37}},\ \bibinfo {pages} {13} (\bibinfo {year}
  {1976})}\BibitemShut {NoStop}%
\bibitem [{\citenamefont {Odijk}(1983)}]{odijk_statistics_1983}%
  \BibitemOpen
  \bibfield  {author} {\bibinfo {author} {\bibfnamefont {T.}~\bibnamefont
  {Odijk}},\ }\bibfield  {title} {\enquote {\bibinfo {title} {The statistics
  and dynamics of confined or entangled stiff polymers},}\ }\href {\doibase
  10.1021/ma00242a015} {\bibfield  {journal} {\bibinfo  {journal}
  {Macromolecules}\ }\textbf {\bibinfo {volume} {16}},\ \bibinfo {pages}
  {1340--1344} (\bibinfo {year} {1983})}\BibitemShut {NoStop}%
\bibitem [{\citenamefont {Semenov}(1986)}]{semenov_dynamics_1986}%
  \BibitemOpen
  \bibfield  {author} {\bibinfo {author} {\bibfnamefont {A.~N.}\ \bibnamefont
  {Semenov}},\ }\bibfield  {title} {\enquote {\bibinfo {title} {Dynamics of
  concentrated solutions of rigid-chain polymers. {Part} 1. {Brownian} motion
  of persistent macromolecules in isotropic solution},}\ }\href {\doibase
  10.1039/F29868200317} {\bibfield  {journal} {\bibinfo  {journal} {Journal of
  the Chemical Society, Faraday Transactions 2: Molecular and Chemical
  Physics}\ }\textbf {\bibinfo {volume} {82}},\ \bibinfo {pages} {317--329}
  (\bibinfo {year} {1986})}\BibitemShut {NoStop}%
\bibitem [{\citenamefont {Morse}(2001)}]{morse_tube_2001}%
  \BibitemOpen
  \bibfield  {author} {\bibinfo {author} {\bibfnamefont {D.~C.}\ \bibnamefont
  {Morse}},\ }\bibfield  {title} {\enquote {\bibinfo {title} {Tube diameter in
  tightly entangled solutions of semiflexible polymers},}\ }\href {\doibase
  10.1103/PhysRevE.63.031502} {\bibfield  {journal} {\bibinfo  {journal}
  {Physical Review E}\ }\textbf {\bibinfo {volume} {63}},\ \bibinfo {pages}
  {031502} (\bibinfo {year} {2001})}\BibitemShut {NoStop}%
\bibitem [{\citenamefont {Tassieri}(2017)}]{tassieri_dynamics_2017}%
  \BibitemOpen
  \bibfield  {author} {\bibinfo {author} {\bibfnamefont {M.}~\bibnamefont
  {Tassieri}},\ }\bibfield  {title} {\enquote {\bibinfo {title} {Dynamics of
  {Semiflexible} {Polymer} {Solutions} in the {Tightly} {Entangled}
  {Concentration} {Regime}},}\ }\href {\doibase 10.1021/acs.macromol.7b01024}
  {\bibfield  {journal} {\bibinfo  {journal} {Macromolecules}\ }\textbf
  {\bibinfo {volume} {50}},\ \bibinfo {pages} {5611--5618} (\bibinfo {year}
  {2017})}\BibitemShut {NoStop}%
\bibitem [{\citenamefont {Burkhardt}(1995)}]{burkhardt_free_1995}%
  \BibitemOpen
  \bibfield  {author} {\bibinfo {author} {\bibfnamefont {T.~W.}\ \bibnamefont
  {Burkhardt}},\ }\bibfield  {title} {\enquote {\bibinfo {title} {Free energy
  of a semiflexible polymer confined along an axis},}\ }\href {\doibase
  10.1088/0305-4470/28/24/001} {\bibfield  {journal} {\bibinfo  {journal}
  {Journal of Physics A: Mathematical and General}\ }\textbf {\bibinfo {volume}
  {28}},\ \bibinfo {pages} {L629--L635} (\bibinfo {year} {1995})}\BibitemShut
  {NoStop}%
\bibitem [{\citenamefont {Schiffels}\ \emph {et~al.}(2013)\citenamefont
  {Schiffels}, \citenamefont {Liedl},\ and\ \citenamefont
  {Fygenson}}]{schiffels_nanoscale_2013}%
  \BibitemOpen
  \bibfield  {author} {\bibinfo {author} {\bibfnamefont {D.}~\bibnamefont
  {Schiffels}}, \bibinfo {author} {\bibfnamefont {T.}~\bibnamefont {Liedl}}, \
  and\ \bibinfo {author} {\bibfnamefont {D.~K.}\ \bibnamefont {Fygenson}},\
  }\bibfield  {title} {\enquote {\bibinfo {title} {Nanoscale {Structure} and
  {Microscale} {Stiffness} of {DNA} {Nanotubes}},}\ }\href {\doibase
  10.1021/nn401362p} {\bibfield  {journal} {\bibinfo  {journal} {ACS Nano}\
  }\textbf {\bibinfo {volume} {7}},\ \bibinfo {pages} {6700--6710} (\bibinfo
  {year} {2013})}\BibitemShut {NoStop}%
\bibitem [{\citenamefont {Glaser}\ \emph {et~al.}(2016)\citenamefont {Glaser},
  \citenamefont {Schnau\ss}, \citenamefont {Tschirner}, \citenamefont
  {Schmidt}, \citenamefont {Moebius-Winkler}, \citenamefont {K\"as},\ and\
  \citenamefont {Smith}}]{glaser_self_2016}%
  \BibitemOpen
  \bibfield  {author} {\bibinfo {author} {\bibfnamefont {M.}~\bibnamefont
  {Glaser}}, \bibinfo {author} {\bibfnamefont {J.}~\bibnamefont {Schnau\ss}},
  \bibinfo {author} {\bibfnamefont {T.}~\bibnamefont {Tschirner}}, \bibinfo
  {author} {\bibfnamefont {B.~U.~S.}\ \bibnamefont {Schmidt}}, \bibinfo
  {author} {\bibfnamefont {M.}~\bibnamefont {Moebius-Winkler}}, \bibinfo
  {author} {\bibfnamefont {J.~A.}\ \bibnamefont {K\"as}}, \ and\ \bibinfo
  {author} {\bibfnamefont {D.~M.}\ \bibnamefont {Smith}},\ }\bibfield  {title}
  {\enquote {\bibinfo {title} {Self-assembly of hierarchically ordered
  structures in {DNA} nanotube systems},}\ }\href {\doibase
  10.1088/1367-2630/18/5/055001} {\bibfield  {journal} {\bibinfo  {journal}
  {New Journal of Physics}\ }\textbf {\bibinfo {volume} {18}},\ \bibinfo
  {pages} {055001} (\bibinfo {year} {2016})}\BibitemShut {NoStop}%
\bibitem [{\citenamefont {Lorenz}\ \emph {et~al.}(2018)\citenamefont {Lorenz},
  \citenamefont {Schnau\ss}, \citenamefont {Glaser}, \citenamefont
  {Sajfutdinow}, \citenamefont {Schuldt}, \citenamefont {K\"as},\ and\
  \citenamefont {Smith}}]{lorenz_synthetic_2018}%
  \BibitemOpen
  \bibfield  {author} {\bibinfo {author} {\bibfnamefont {J.~S.}\ \bibnamefont
  {Lorenz}}, \bibinfo {author} {\bibfnamefont {J.}~\bibnamefont {Schnau\ss}},
  \bibinfo {author} {\bibfnamefont {M.}~\bibnamefont {Glaser}}, \bibinfo
  {author} {\bibfnamefont {M.}~\bibnamefont {Sajfutdinow}}, \bibinfo {author}
  {\bibfnamefont {C.}~\bibnamefont {Schuldt}}, \bibinfo {author} {\bibfnamefont
  {J.~A.}\ \bibnamefont {K\"as}}, \ and\ \bibinfo {author} {\bibfnamefont
  {D.~M.}\ \bibnamefont {Smith}},\ }\bibfield  {title} {\enquote {\bibinfo
  {title} {Synthetic {Transient} {Crosslinks} {Program} the {Mechanics} of
  {Soft}, {Biopolymer}-{Based} {Materials}},}\ }\href {\doibase
  10.1002/adma.201706092} {\bibfield  {journal} {\bibinfo  {journal} {Advanced
  Materials}\ }\textbf {\bibinfo {volume} {30}},\ \bibinfo {pages} {1706092}
  (\bibinfo {year} {2018})}\BibitemShut {NoStop}%
\bibitem{supplemental} See Supplemental Material.
\bibitem [{\citenamefont {Isambert}\ \emph {et~al.}(1995)\citenamefont
  {Isambert}, \citenamefont {Venier}, \citenamefont {Maggs}, \citenamefont
  {Fattoum}, \citenamefont {Kassab}, \citenamefont {Pantaloni},\ and\
  \citenamefont {Carlier}}]{isambert_flexibility_1995}%
  \BibitemOpen
  \bibfield  {author} {\bibinfo {author} {\bibfnamefont {H.}~\bibnamefont
  {Isambert}}, \bibinfo {author} {\bibfnamefont {P.}~\bibnamefont {Venier}},
  \bibinfo {author} {\bibfnamefont {A.~C.}\ \bibnamefont {Maggs}}, \bibinfo
  {author} {\bibfnamefont {A.}~\bibnamefont {Fattoum}}, \bibinfo {author}
  {\bibfnamefont {R.}~\bibnamefont {Kassab}}, \bibinfo {author} {\bibfnamefont
  {D.}~\bibnamefont {Pantaloni}}, \ and\ \bibinfo {author} {\bibfnamefont
  {M.~F.}\ \bibnamefont {Carlier}},\ }\bibfield  {title} {\enquote {\bibinfo
  {title} {Flexibility of actin filaments derived from thermal fluctuations.
  {Effect} of bound nucleotide, phalloidin, and muscle regulatory proteins},}\
  }\href {\doibase 10.1074/jbc.270.19.11437} {\bibfield  {journal} {\bibinfo
  {journal} {The Journal of Biological Chemistry}\ }\textbf {\bibinfo {volume}
  {270}},\ \bibinfo {pages} {11437--11444} (\bibinfo {year}
  {1995})}\BibitemShut {NoStop}%
\bibitem [{\citenamefont {Smith}\ \emph {et~al.}(2010)\citenamefont {Smith},
  \citenamefont {Li}, \citenamefont {Shen}, \citenamefont {Huang},
  \citenamefont {Yusuf},\ and\ \citenamefont
  {Vavylonis}}]{smith_segmentation_2010}%
  \BibitemOpen
  \bibfield  {author} {\bibinfo {author} {\bibfnamefont {M.~B.}\ \bibnamefont
  {Smith}}, \bibinfo {author} {\bibfnamefont {H.}~\bibnamefont {Li}}, \bibinfo
  {author} {\bibfnamefont {T.}~\bibnamefont {Shen}}, \bibinfo {author}
  {\bibfnamefont {X.}~\bibnamefont {Huang}}, \bibinfo {author} {\bibfnamefont
  {E.}~\bibnamefont {Yusuf}}, \ and\ \bibinfo {author} {\bibfnamefont
  {D.}~\bibnamefont {Vavylonis}},\ }\bibfield  {title} {\enquote {\bibinfo
  {title} {Segmentation and {Tracking} of {Cytoskeletal} {Filaments} {Using}
  {Open} {Active} {Contours}},}\ }\href {\doibase 10.1002/cm.20481} {\bibfield
  {journal} {\bibinfo  {journal} {Cytoskeleton}\ }\textbf {\bibinfo {volume}
  {67}},\ \bibinfo {pages} {693--705} (\bibinfo {year} {2010})}\BibitemShut
  {NoStop}%
\bibitem [{\citenamefont {Marrucci}(1985)}]{marrucci_relaxation_1985}%
  \BibitemOpen
  \bibfield  {author} {\bibinfo {author} {\bibfnamefont {G.}~\bibnamefont
  {Marrucci}},\ }\bibfield  {title} {\enquote {\bibinfo {title} {Relaxation by
  reptation and tube enlargement: {A} model for polydisperse polymers},}\
  }\href {\doibase 10.1002/pol.1985.180230115} {\bibfield  {journal} {\bibinfo
  {journal} {Journal of Polymer Science: Polymer Physics Edition}\ }\textbf
  {\bibinfo {volume} {23}},\ \bibinfo {pages} {159--177} (\bibinfo {year}
  {1985})}\BibitemShut {NoStop}%
\bibitem [{\citenamefont {Golde}\ \emph {et~al.}(2018)\citenamefont {Golde},
  \citenamefont {Huster}, \citenamefont {Glaser}, \citenamefont {H\"andler},
  \citenamefont {Herrmann}, \citenamefont {K\"as},\ and\ \citenamefont
  {Schnau\ss}}]{golde_glassy_2018}%
  \BibitemOpen
  \bibfield  {author} {\bibinfo {author} {\bibfnamefont {T.}~\bibnamefont
  {Golde}}, \bibinfo {author} {\bibfnamefont {C.}~\bibnamefont {Huster}},
  \bibinfo {author} {\bibfnamefont {M.}~\bibnamefont {Glaser}}, \bibinfo
  {author} {\bibfnamefont {T.}~\bibnamefont {H\"andler}}, \bibinfo {author}
  {\bibfnamefont {H.}~\bibnamefont {Herrmann}}, \bibinfo {author}
  {\bibfnamefont {J.~A.}\ \bibnamefont {K\"as}}, \ and\ \bibinfo {author}
  {\bibfnamefont {J.}~\bibnamefont {Schnau\ss}},\ }\bibfield  {title} {\enquote
  {\bibinfo {title} {Glassy dynamics in composite biopolymer networks},}\
  }\href {\doibase 10.1039/C8SM01061G} {\bibfield  {journal} {\bibinfo
  {journal} {Soft Matter}\ }\textbf {\bibinfo {volume} {14}},\ \bibinfo {pages}
  {7970--7978} (\bibinfo {year} {2018})}\BibitemShut {NoStop}%
\bibitem [{\citenamefont {Schmidt}\ \emph {et~al.}(1989)\citenamefont
  {Schmidt}, \citenamefont {Baermann}, \citenamefont {Isenberg},\ and\
  \citenamefont {Sackmann}}]{schmidt_chain_1989}%
  \BibitemOpen
  \bibfield  {author} {\bibinfo {author} {\bibfnamefont {C.~F.}\ \bibnamefont
  {Schmidt}}, \bibinfo {author} {\bibfnamefont {M.}~\bibnamefont {Baermann}},
  \bibinfo {author} {\bibfnamefont {G.}~\bibnamefont {Isenberg}}, \ and\
  \bibinfo {author} {\bibfnamefont {E.}~\bibnamefont {Sackmann}},\ }\bibfield
  {title} {\enquote {\bibinfo {title} {Chain dynamics, mesh size, and diffusive
  transport in networks of polymerized actin: a quasielastic light scattering
  and microfluorescence study},}\ }\href {\doibase 10.1021/ma00199a023}
  {\bibfield  {journal} {\bibinfo  {journal} {Macromolecules}\ }\textbf
  {\bibinfo {volume} {22}},\ \bibinfo {pages} {3638--3649} (\bibinfo {year}
  {1989})}\BibitemShut {NoStop}%
\bibitem [{\citenamefont {Wang}\ \emph {et~al.}(2010)\citenamefont {Wang},
  \citenamefont {Guan}, \citenamefont {Anthony}, \citenamefont {Bae},
  \citenamefont {Schweizer},\ and\ \citenamefont
  {Granick}}]{wang_confining_2010}%
  \BibitemOpen
  \bibfield  {author} {\bibinfo {author} {\bibfnamefont {B.}~\bibnamefont
  {Wang}}, \bibinfo {author} {\bibfnamefont {J.}~\bibnamefont {Guan}}, \bibinfo
  {author} {\bibfnamefont {S.~M.}\ \bibnamefont {Anthony}}, \bibinfo {author}
  {\bibfnamefont {S.~C.}\ \bibnamefont {Bae}}, \bibinfo {author} {\bibfnamefont
  {Ke.~S.}\ \bibnamefont {Schweizer}}, \ and\ \bibinfo {author} {\bibfnamefont
  {S.}~\bibnamefont {Granick}},\ }\bibfield  {title} {\enquote {\bibinfo
  {title} {Confining potential when a biopolymer filament reptates},}\ }\href
  {\doibase 10.1103/PhysRevLett.104.118301} {\bibfield  {journal} {\bibinfo
  {journal} {Phys. Rev. Lett.}\ }\textbf {\bibinfo {volume} {104}},\ \bibinfo
  {pages} {118301} (\bibinfo {year} {2010})}\BibitemShut {NoStop}%
\bibitem [{\citenamefont {Cherstvy}(2011)}]{cherstvy_electrostatic_2011}%
  \BibitemOpen
  \bibfield  {author} {\bibinfo {author} {\bibfnamefont {A.~G.}\ \bibnamefont
  {Cherstvy}},\ }\bibfield  {title} {\enquote {\bibinfo {title} {Electrostatic
  interactions in biological {DNA}-related systems},}\ }\href {\doibase
  10.1039/C0CP02796K} {\bibfield  {journal} {\bibinfo  {journal} {Phys. Chem.
  Chem. Phys.}\ }\textbf {\bibinfo {volume} {13}},\ \bibinfo {pages}
  {9942--9968} (\bibinfo {year} {2011})}\BibitemShut {NoStop}%
\bibitem [{\citenamefont {Golde}\ \emph {et~al.}(2019)\citenamefont {Golde},
  \citenamefont {Glaser}, \citenamefont {Tutmarc}, \citenamefont {Elbalasy},
  \citenamefont {Huster}, \citenamefont {Busteros}, \citenamefont {Smith},
  \citenamefont {Herrmann}, \citenamefont {K\"as},\ and\ \citenamefont
  {Schnau\ss}}]{golde_role_2019}%
  \BibitemOpen
  \bibfield  {author} {\bibinfo {author} {\bibfnamefont {T.}~\bibnamefont
  {Golde}}, \bibinfo {author} {\bibfnamefont {M.}~\bibnamefont {Glaser}},
  \bibinfo {author} {\bibfnamefont {C.}~\bibnamefont {Tutmarc}}, \bibinfo
  {author} {\bibfnamefont {I.}~\bibnamefont {Elbalasy}}, \bibinfo {author}
  {\bibfnamefont {C.}~\bibnamefont {Huster}}, \bibinfo {author} {\bibfnamefont
  {G.}~\bibnamefont {Busteros}}, \bibinfo {author} {\bibfnamefont {D.~M.}\
  \bibnamefont {Smith}}, \bibinfo {author} {\bibfnamefont {H.}~\bibnamefont
  {Herrmann}}, \bibinfo {author} {\bibfnamefont {J.~A.}\ \bibnamefont {K\"as}},
  \ and\ \bibinfo {author} {\bibfnamefont {J.}~\bibnamefont {Schnau\ss}},\
  }\bibfield  {title} {\enquote {\bibinfo {title} {The role of stickiness in
  the rheology of semiflexible polymers},}\ }\href {\doibase
  10.1039/C9SM00433E} {\bibfield  {journal} {\bibinfo  {journal} {Soft Matter}\
  }\textbf {\bibinfo {volume} {15}},\ \bibinfo {pages} {4865--4872} (\bibinfo
  {year} {2019})}\BibitemShut {NoStop}%
\bibitem [{\citenamefont {Glaser}\ \emph {et~al.}(2010)\citenamefont {Glaser},
  \citenamefont {Chakraborty}, \citenamefont {Kroy}, \citenamefont {Lauter},
  \citenamefont {Degawa}, \citenamefont {Kirchge{\ss}ner}, \citenamefont
  {Hoffmann}, \citenamefont {Merkel},\ and\ \citenamefont
  {Giesen}}]{glaser_tube_2010}%
  \BibitemOpen
  \bibfield  {author} {\bibinfo {author} {\bibfnamefont {J.}~\bibnamefont
  {Glaser}}, \bibinfo {author} {\bibfnamefont {D.}~\bibnamefont {Chakraborty}},
  \bibinfo {author} {\bibfnamefont {K.}~\bibnamefont {Kroy}}, \bibinfo {author}
  {\bibfnamefont {I.}~\bibnamefont {Lauter}}, \bibinfo {author} {\bibfnamefont
  {M.}~\bibnamefont {Degawa}}, \bibinfo {author} {\bibfnamefont
  {N.}~\bibnamefont {Kirchge{\ss}ner}}, \bibinfo {author} {\bibfnamefont
  {B.}~\bibnamefont {Hoffmann}}, \bibinfo {author} {\bibfnamefont
  {R.}~\bibnamefont {Merkel}}, \ and\ \bibinfo {author} {\bibfnamefont
  {M.}~\bibnamefont {Giesen}},\ }\bibfield  {title} {\enquote {\bibinfo {title}
  {Tube {Width} {Fluctuations} in {F}-{Actin} {Solutions}},}\ }\href {\doibase
  10.1103/PhysRevLett.105.037801} {\bibfield  {journal} {\bibinfo  {journal}
  {Physical Review Letters}\ }\textbf {\bibinfo {volume} {105}},\ \bibinfo
  {pages} {037801} (\bibinfo {year} {2010})}\BibitemShut {NoStop}%
\bibitem [{\citenamefont {Fakhri}\ \emph {et~al.}(2010)\citenamefont {Fakhri},
  \citenamefont {MacKintosh}, \citenamefont {Lounis}, \citenamefont {Cognet},\
  and\ \citenamefont {Pasquali}}]{fakhri_brownian_2010}%
  \BibitemOpen
  \bibfield  {author} {\bibinfo {author} {\bibfnamefont {N.}~\bibnamefont
  {Fakhri}}, \bibinfo {author} {\bibfnamefont {F.~C.}\ \bibnamefont
  {MacKintosh}}, \bibinfo {author} {\bibfnamefont {B.}~\bibnamefont {Lounis}},
  \bibinfo {author} {\bibfnamefont {L.}~\bibnamefont {Cognet}}, \ and\ \bibinfo
  {author} {\bibfnamefont {M.}~\bibnamefont {Pasquali}},\ }\bibfield  {title}
  {\enquote {\bibinfo {title} {Brownian {Motion} of {Stiff} {Filaments} in a
  {Crowded} {Environment}},}\ }\href {\doibase 10.1126/science.1197321}
  {\bibfield  {journal} {\bibinfo  {journal} {Science}\ }\textbf {\bibinfo
  {volume} {330}},\ \bibinfo {pages} {1804--1807} (\bibinfo {year}
  {2010})}\BibitemShut {NoStop}%
\bibitem [{\citenamefont {Gardel}\ \emph {et~al.}(2004)\citenamefont {Gardel},
  \citenamefont {Shin}, \citenamefont {MacKintosh}, \citenamefont {Mahadevan},
  \citenamefont {Matsudaira},\ and\ \citenamefont
  {Weitz}}]{gardel_elastic_2004}%
  \BibitemOpen
  \bibfield  {author} {\bibinfo {author} {\bibfnamefont {M.~L.}\ \bibnamefont
  {Gardel}}, \bibinfo {author} {\bibfnamefont {J.~H.}\ \bibnamefont {Shin}},
  \bibinfo {author} {\bibfnamefont {F.~C.}\ \bibnamefont {MacKintosh}},
  \bibinfo {author} {\bibfnamefont {L.}~\bibnamefont {Mahadevan}}, \bibinfo
  {author} {\bibfnamefont {P.}~\bibnamefont {Matsudaira}}, \ and\ \bibinfo
  {author} {\bibfnamefont {D.~A.}\ \bibnamefont {Weitz}},\ }\bibfield  {title}
  {\enquote {\bibinfo {title} {Elastic {Behavior} of {Cross}-{Linked} and
  {Bundled} {Actin} {Networks}},}\ }\href {\doibase 10.1126/science.1095087}
  {\bibfield  {journal} {\bibinfo  {journal} {Science}\ }\textbf {\bibinfo
  {volume} {304}},\ \bibinfo {pages} {1301--1305} (\bibinfo {year}
  {2004})}\BibitemShut {NoStop}%
\bibitem [{\citenamefont {Lang}\ and\ \citenamefont
  {Frey}(2018)}]{lang_disentangling_2018}%
  \BibitemOpen
  \bibfield  {author} {\bibinfo {author} {\bibfnamefont {P.}~\bibnamefont
  {Lang}}\ and\ \bibinfo {author} {\bibfnamefont {E.}~\bibnamefont {Frey}},\
  }\bibfield  {title} {\enquote {\bibinfo {title} {Disentangling entanglements
  in biopolymer solutions},}\ }\href {\doibase 10.1038/s41467-018-02837-5}
  {\bibfield  {journal} {\bibinfo  {journal} {Nature Communications}\ }\textbf
  {\bibinfo {volume} {9}},\ \bibinfo {pages} {494} (\bibinfo {year}
  {2018})}\BibitemShut {NoStop}%
\bibitem [{\citenamefont {Golde}\ \emph {et~al.}(2013)\citenamefont {Golde},
  \citenamefont {Schuldt}, \citenamefont {Schnau\ss}, \citenamefont {Strehle},
  \citenamefont {Glaser},\ and\ \citenamefont
  {K\"as}}]{golde_fluorescent_2013}%
  \BibitemOpen
  \bibfield  {author} {\bibinfo {author} {\bibfnamefont {T.}~\bibnamefont
  {Golde}}, \bibinfo {author} {\bibfnamefont {C.}~\bibnamefont {Schuldt}},
  \bibinfo {author} {\bibfnamefont {J.}~\bibnamefont {Schnau\ss}}, \bibinfo
  {author} {\bibfnamefont {D.}~\bibnamefont {Strehle}}, \bibinfo {author}
  {\bibfnamefont {M.}~\bibnamefont {Glaser}}, \ and\ \bibinfo {author}
  {\bibfnamefont {J.}~\bibnamefont {K\"as}},\ }\bibfield  {title} {\enquote
  {\bibinfo {title} {Fluorescent beads disintegrate actin networks},}\ }\href
  {\doibase 10.1103/PhysRevE.88.044601} {\bibfield  {journal} {\bibinfo
  {journal} {Physical Review E}\ }\textbf {\bibinfo {volume} {88}},\ \bibinfo
  {pages} {044601} (\bibinfo {year} {2013})}\BibitemShut {NoStop}%
\bibitem [{\citenamefont {Fischer}\ \emph {et~al.}(2019)\citenamefont
  {Fischer}, \citenamefont {Hayn},\ and\ \citenamefont
  {Mierke}}]{fischer_fast_2019}%
  \BibitemOpen
  \bibfield  {author} {\bibinfo {author} {\bibfnamefont {T.}~\bibnamefont
  {Fischer}}, \bibinfo {author} {\bibfnamefont {A.}~\bibnamefont {Hayn}}, \
  and\ \bibinfo {author} {\bibfnamefont {C.~T.}\ \bibnamefont {Mierke}},\
  }\bibfield  {title} {\enquote {\bibinfo {title} {Fast and reliable advanced
  two-step pore-size analysis of biomimetic {3D} extracellular matrix
  scaffolds},}\ }\href {\doibase 10.1038/s41598-019-44764-5} {\bibfield
  {journal} {\bibinfo  {journal} {Scientific Reports}\ }\textbf {\bibinfo
  {volume} {9}},\ \bibinfo {pages} {8352} (\bibinfo {year} {2019})}\BibitemShut
  {NoStop}%
\end{thebibliography}
\end{document}


\title{Supplemental Material: Measuring structural parameters of crosslinked and entangled semiflexible polymer networks with single-filament tracing}

\author{Tina Händler}
\affiliation{Peter Debye Institute for Soft Matter Physics, Universität Leipzig, Linnéstraße 5, 04103 Leipzig, Germany}
\affiliation{Fraunhofer Institute for Cell Therapy and Immunology, Perlickstraße 1, 04103 Leipzig, Germany}
\author{Cary Tutmarc}
\affiliation{Peter Debye Institute for Soft Matter Physics, Universität Leipzig, Linnéstraße 5, 04103 Leipzig, Germany}
\affiliation{Fraunhofer Institute for Cell Therapy and Immunology, Perlickstraße 1, 04103 Leipzig, Germany}
\author{Martin Glaser}
\affiliation{Peter Debye Institute for Soft Matter Physics, Universität Leipzig, Linnéstraße 5, 04103 Leipzig, Germany}
\affiliation{Fraunhofer Institute for Cell Therapy and Immunology, Perlickstraße 1, 04103 Leipzig, Germany}
\author{Jessica S. Freitag}
\affiliation{Fraunhofer Institute for Cell Therapy and Immunology, Perlickstraße 1, 04103 Leipzig, Germany}
\author{David M. Smith}
\affiliation{Fraunhofer Institute for Cell Therapy and Immunology, Perlickstraße 1, 04103 Leipzig, Germany}
\affiliation{Institute of Clinical Immunology, University of Leipzig Medical Faculty, 04103 Leipzig, Germany}
\affiliation{Dhirubhai Ambani Institute of Information and Communication Technology, Gandhinagar 382 007, India}
\author{Jörg Schnauß}
\affiliation{Peter Debye Institute for Soft Matter Physics, Universität Leipzig, Linnéstraße 5, 04103 Leipzig, Germany}
\affiliation{Fraunhofer Institute for Cell Therapy and Immunology, Perlickstraße 1, 04103 Leipzig, Germany}
\affiliation{Unconventional Computing Laboratory, Department of Computer Science, University of the West of England, Bristol, UK} 
\date{\today}

\maketitle

\section{Material and Methods}

\subsection{DNA $n$-helix tubes ($n$HT)}
Lyophilized oligonucleotides (biomers.net GmbH, Germany) were resuspended in millipore water (sequences are given in \cite{schuldt_tuning_2016} and \cite{yin_programming_2008}) according to the manufacturer.
Oligonucleotide concentration was determined using a NanoDrop 1000 (Thermo Fisher Scientific Inc., USA).
Final DNA \(n\)HT samples consisted of the $n-1$ strands $\text{U}_1 - \text{U}_{n-1}$ and one $\text{T}_n$ strand, each at the same monomer concentration.
Final buffer conditions for hybridization were 1xTE (\SI{10}{\milli\molar} Tris, \SI{1}{\milli\molar} EDTA, pH \num{8}) and \SI{12.5}{\milli\molar} \ce{MgCl2}.
A thermocycler (TProfessional Standard PCR Thermocycler, Core Life Sciences Inc., USA) was used to hybridize the \(n\)HTs. The hybridization protocol included randomization and dehybridization for \SI{10}{\minute} at \SI{90}{\celsius} followed by complementary base pairing in 20 temperature steps of \SI{-0.5}{\celsius} for \SI{60}{\minute} each, starting from \SI{65}{\celsius} and a quick drop to \SI{20}{\celsius}.
Hybridized \(n\)HTs were stored for up to 4 weeks at room temperature without detectable degradation.
For fluorescently labeled $n$HTs as tracer filaments, $\text{U}_1$ was substituted by the modified $\text{U}_1$-Cy3, with the dye Cy3 attached to the regular $\text{U}_1$ strand.

\subsection{Actin}
G-actin was prepared from rabbit muscle as described previously \cite{gentry_buckling-induced_2009}. Actin polymerization was initiated by the addition of 10 times concentrated F-Buffer (\SI{1}{\molar} KCl, \SI{10}{\milli\molar} \ce{MgCl2}, \SI{2}{\milli\molar} ATP, \SI{10}{\milli\molar} DTT, \SI{20}{\milli\molar} sodium phosphate, pH 7.5).

\subsection{Crosslinker wLX}

The crosslinker wLX was synthesized as described by \citet{lorenz_synthetic_2018}. In short, two actin-binding domains (LifeAct\textsuperscript{\textregistered}, Peptide Specialty Laboratories GmbH, Germany) were covalently attached to the ends of double-stranded DNA (sequences given in \cite{lorenz_synthetic_2018}) using a copper-free click-chemistry approach. Since the two LifeAct\textsuperscript{\textregistered} peptides were connected via double-stranded DNA, the composition physically crosslinked actin filaments.

\subsection{Sample preparation and measurement}

Pre-hybridized \(n\)HTs labeled with the dye Cy3 were diluted gradually to \SI{10}{\nano\molar}.
From this dilution, a small fraction was gently mixed with a pre-hybridized unlabeled 8HT network to a final ratio of 1:4000 of labeled to unlabeled filaments. Subsequently the sample solution was placed between two glass slides previously coated with Sigmacote (Sigma-Aldrich, USA) and sealed with grease and nail polish. 8HT network samples were equilibrated overnight at \SI{4}{\celsius} and measured at \SI{8}{\micro\molar}.\\
For the actin samples, Cy3-labeled nanotubes were mixed with monomeric unlabeled actin (and additionally with the crosslinker wLX to a molar ratio of 150 actin monomers to 1 crosslinker molecule for the crosslinked actin networks) to an end ratio of nanotubes to actin filaments of at least 1:6000. Actin assembly was triggered by addition of 10x F-Buffer and actin filaments were allowed to polymerize for \SIrange{1}{2}{\hour}, ensuring that the actin network formed around the embedded tracer filaments. The samples were then placed between two glass slides previously passivated with \SI{5}{\percent} bovine serum albumin or Sigmacote (Sigma-Aldrich, USA), sealed with grease and left to settle for \SIrange{30}{60}{\minute} prior to measurement. Entangled actin networks were polymerized at \SI{0.5}{\milli\gram\per\milli\litre} and crosslinked networks at \SIrange{0.3}{0.9}{\milli\gram\per\milli\litre}.\\
After the equilibration time, images of labeled reptating \(n\)HT tubes were recorded via an epi-fluorescence microscope (Leica DM-IRB, 100x objective, NA 1.35) with an attached CCD camera (Andor, iXon DV887) at a frame rate of \SI{10}{\hertz}.

\subsection{Tubewidth analysis}

\begin{figure*}
  \centering
  \includegraphics[width=\textwidth]{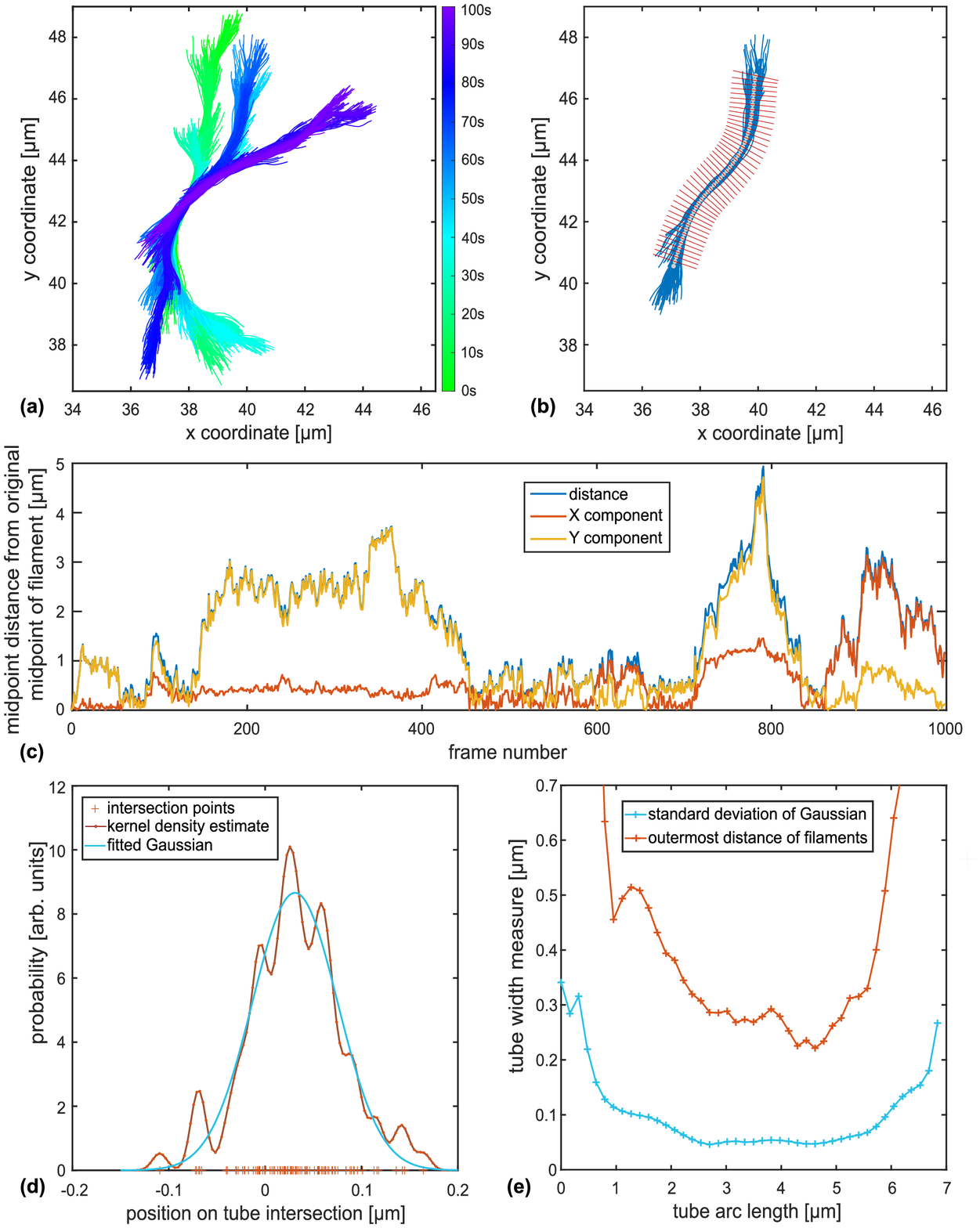}
  \caption{Scheme of tube width determination, example filament is an 8HT in F-actin at \SI{0.5}{\milli\gram\per\milli\litre} crosslinked with wLX at a ratio of 150:1. The process is explained in detail in the text.} 
	\label{fig_tw_det}
\end{figure*}

After preprocessing the image stacks with the ImageJ plugin Stack Contrast Adjustment \cite{capek_methods_2006}, filament backbones were determined by the ImageJ plugin JFilament \cite{smith_segmentation_2010}. From the backbones, tube widths were derived using a self-written Matlab (The MathWorks, Inc.) script as depicted by Fig. \ref{fig_tw_det}. The process is as follows: Panel \ref{fig_tw_det}(a) plots all 1000 tracked configurations of the filament during the observation time of \SI{100}{\second} (see color bar at side). From these 1000 frames, three sets of 100 consecutive frames are chosen so that the filament stays in one reptation tube for the selected \SI{10}{\second}. As a criterion, the deviation of the filament's midpoint from the reference midpoint of the first frame is examined as displayed in panel \ref{fig_tw_det}(c). Subfigure \ref{fig_tw_det}(b) shows an example of one chosen set of 100 tracked configurations (blue). The grey dotted line depicts the mean of all 100 configurations and the red lines are perpendicular to this tube backbone. To determine the tube width, the intersection points of the 100 individual configurations with the orthogonal lines are detected. The resulting intersection positions (orange crosses) are processed for each tube backbone point as shown in panel \ref{fig_tw_det}(d): From the intersection positions, a kernel density distribution is calculated (red). A Gaussian distribution (blue) is fitted to this kernel density estimate. Since this process is repeated for every orthogonal line from panel \ref{fig_tw_det}(b), we can plot a tube width measure for each point of the tube backbone as in subfigure \ref{fig_tw_det}(e). The orange curve shows the distance of the two outermost individual configurations, whereas the blue line is the standard deviation of the Gaussian fit from panel \ref{fig_tw_det}(d). Due to a rather high fluctuation of the filament's ends, only the middle part of the curve in panel \ref{fig_tw_det}(e) is used (i.e., tube arc length from \SI{1}{\micro\meter} to \SI{5.6}{\micro\meter} for this example) and the resulting tube width of the chosen set is the average of two times the standard deviation of the Gaussian fit. The tube width of the whole filament is defined as the average of the tube width from the three subsets of configurations, the standard error of the tube width \(u(a)\) is the standard error of this mean. To implicitly calculate the mesh size from
\begin{equation}
a \approx 0.31\frac{\xi^{6/5}}{l_{\text{p}}^{1/5}} + 0.59\frac{\xi^2}{L},
\label{SIeq_xi_hinsch}
\end{equation}
the contour length of the filament was deduced as the average contour length of the three hundred frames used for tube width determination. Analogously to the tube width, the standard error \(u(L)\) of the contour length is the standard error of this mean.
For the mesh size, we calculated the error \(u(\xi)\) by relative error propagation using 
\(u(\xi) = u(l_{\text{p}})\,\epsilon\,5/6+u(a)5/6+u(L)/2\), with \(u(l_{\text{p}})\) being the standard error of the persistence length according to Table \ref{tab_lengths}. The factor \(\epsilon\) is the exponent extracted from the fit displayed in Fig. 2 in the main text, i.e., \(\epsilon=0.33\) for entangled F-actin, \(\epsilon=0.42\) for crosslinked F-actin and \(\epsilon=0.74\) for the 8HT background network.

\section{Characteristic lengths of DNA nanotube tracers}

In Table \ref{tab_lengths}, we give an overview of the \(n\)HT persistence lengths \(l_{\text{p}}\) and filament diameters \(d_f\) used throughout our calculations. Filament diameters were estimated from \(n\)HT circumference values given in Ref. \cite{yin_programming_2008}. Persistence length values were taken from Ref. \cite{schuldt_tuning_2016}, except for 7HT, where we measured the persistence length with the method described in \cite{schuldt_tuning_2016}. Since the persistence length of actin filaments is about \SI{9}{\micro\meter} \cite{isambert_flexibility_1995}, approximate ratios between tracer (6HT, 7HT, 8HT, 9HT, and 10HT) and background filament (F-actin or 8HT) persistence lengths are 3x, 1.5x, 1x, 1x, and (1/1.5)x, respectively.

\begin{table}[h]
\centering
\begin{tabular}{|c  c  c|} 
 \hline
 tracer & \(l_{\text{p}}\) [\SI{}{\micro\meter}] & \(d_{\text{f}}\) [\SI{}{\nano\meter}]\\
 [0.5ex] 
 \hline
 6HT & 3.18\(\pm\)0.25 & 5.7\\
  7HT & 5.64\(\pm\)0.76 & 6.7\\ 
   8HT & 8.94\(\pm\)0.87 & 7.6\\ 
    9HT & 9.66\(\pm\)1.57 & 8.6\\ 
     10HT & 12.75\(\pm\)1.23 & 9.5\\
     [0.5ex] 
 \hline
\end{tabular}
\caption{Characteristic lengths of the tracer filaments \cite{yin_programming_2008, schuldt_tuning_2016}, \(l_{\text{p}}\) is the persistence length and \(d_{\text{f}}\) is the tracer thickness.}
\label{tab_lengths}
\end{table}

\section{Contour length correction of tube width}

\begin{figure*}
  \centering
  \includegraphics[width=\textwidth]{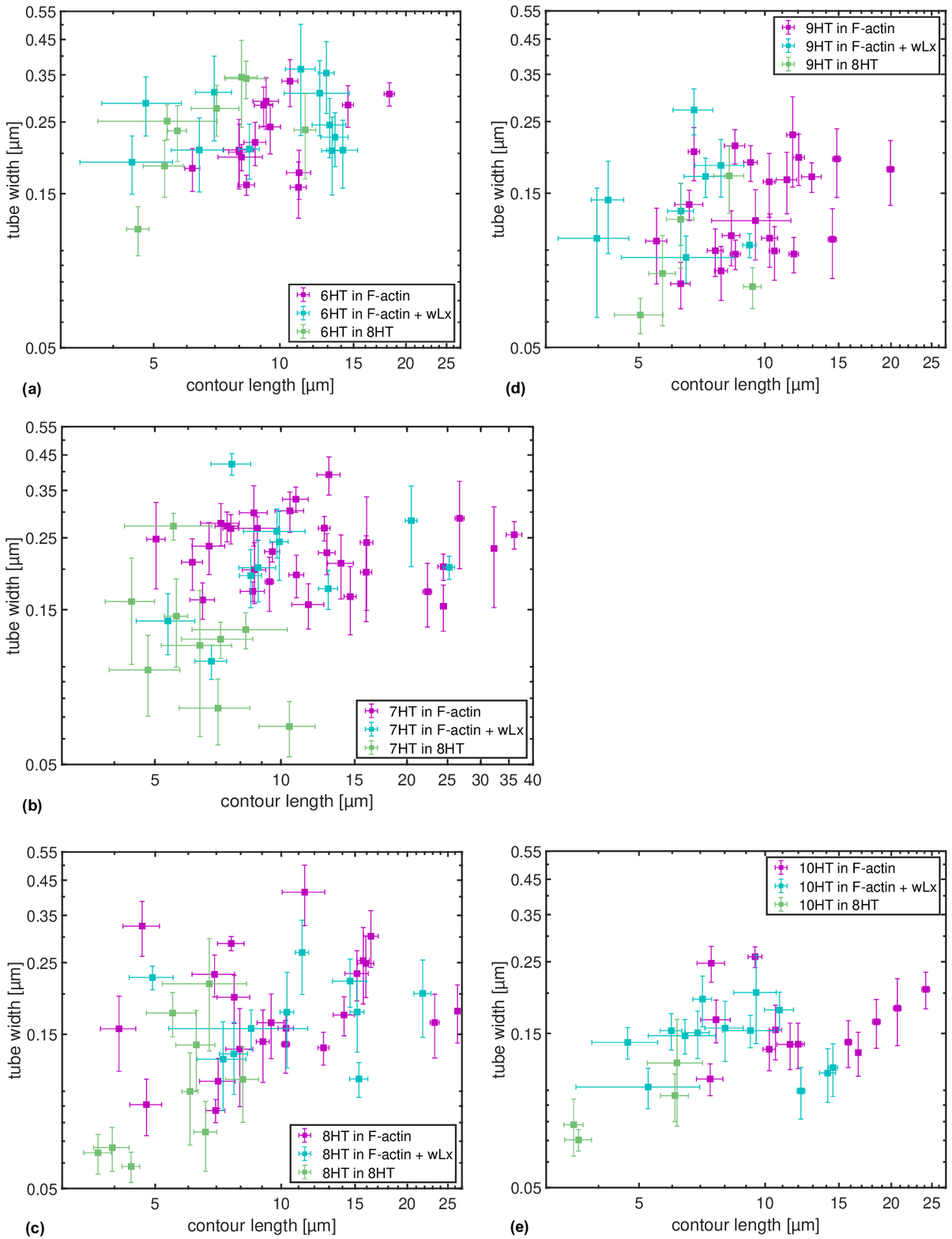}
  \caption{Tube width vs. contour length for all individual tracer filaments which are averaged in Fig. 1 (see main text). The predicted scaling $a \propto L^{-1}$ is not observed for 6HT (a), 7HT (b), 8HT (c), 9HT (d) and 10HT (e) in three different background networks.}
	\label{fig_SItubecontour}
\end{figure*}

\begin{figure*}
  \centering
  \includegraphics[width=\textwidth]{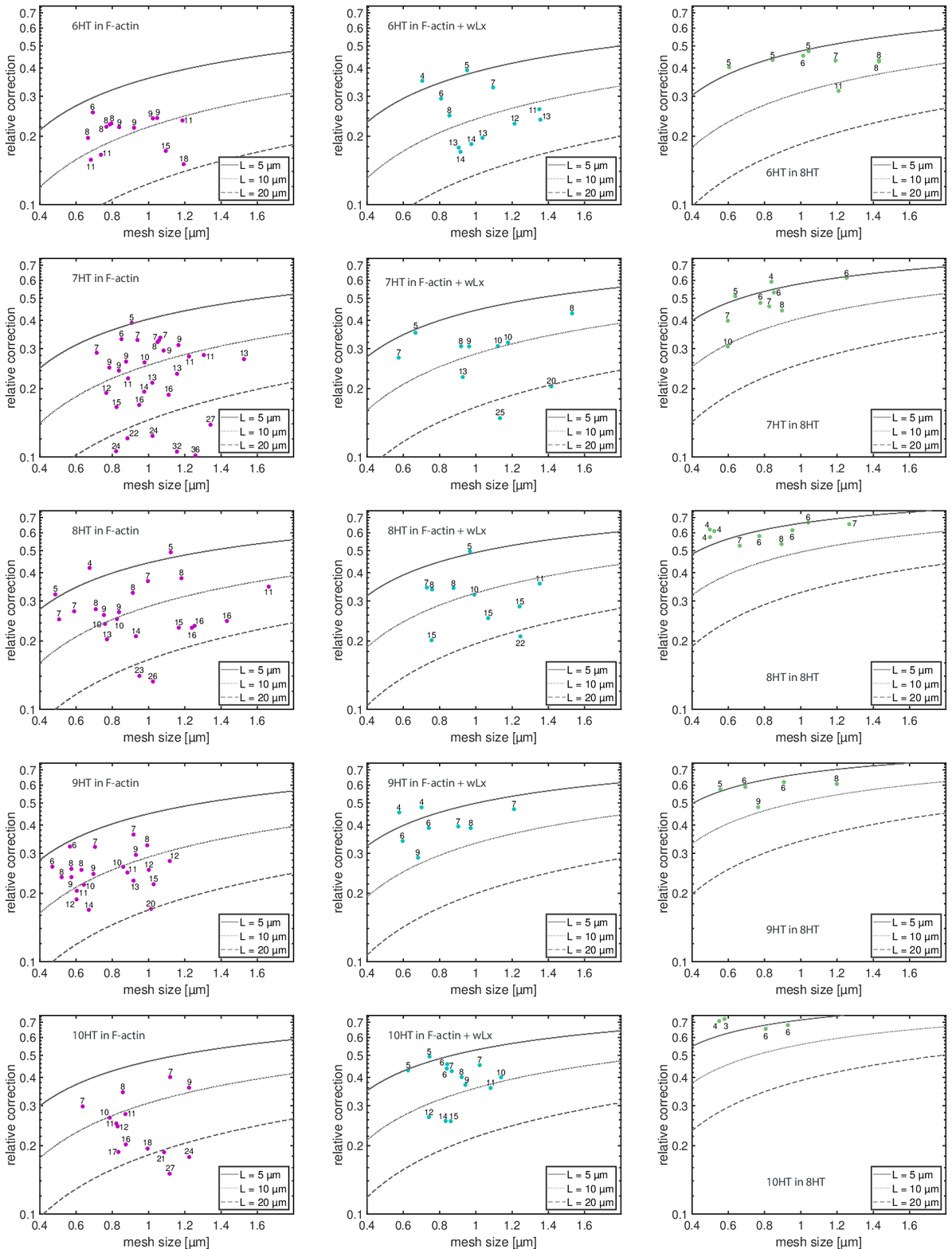}
  \caption{Relative tube width correction vs. calculated mesh size for all data sets of different \(n\)HTs (varying from top to bottom) embedded in three background networks (varying from left to right). The curves for 5, 10 and \SI{20}{\micro\meter} contour length are predictions, while the coloured dots are data points, each accompanied by the rounded contour length of the respective tracer filament.} 
	\label{fig_SIrelcoor}
\end{figure*}

We used Eq. \eqref{SIeq_xi_hinsch} to calculate mesh sizes from measured tube widths. The second term accounts for boundary effects occuring at the end of the reptation tube. In the main text, we showed that the scaling \(a\propto L^{-1}\) is not observed for 8HT tracers embedded in F-actin networks crosslinked with wLx at different concentrations [see Fig. 4(a)]. In Fig. \ref{fig_SItubecontour}, we plot individual tube widths vs. contour lengths of five \(n\)HT tracer types in three different background networks with constant concentrations. The results 
do not correspond to the predicted scaling \(a\propto L^{-1}\) as shown before, again confirming that the variation of the absolute values of tube width due to finite length effects is negligible.
By calculating the relative correction of the tube width stemming from finite length effects, we are able to verify that these effects are small, but larger for shorter tracer filaments [see Fig. 4(b) in main text]. Here we show that the same results are obtained for all measured combinations of tracer filaments and background networks. In Fig. \ref{fig_SIrelcoor}, the relative correction $(0.59\frac{\xi^2}{L})\frac{1}{a}$ is plotted against the mesh size determined by \(a \propto 0.31\frac{\xi^{6/5}}{l_{\text{p}}^{\epsilon}} + 0.59\frac{\xi^2}{L}\) with an exponent \(\epsilon=0.33\) for entangled F-actin, \(\epsilon=0.42\) for crosslinked F-actin and \(\epsilon=0.74\) for the 8HT network, as extracted from the fit displayed in Fig. 2 in the main text.

